\documentclass[aps,prb,twocolumn]{revtex4-1}
\usepackage{graphicx}
\usepackage{dcolumn}
\usepackage{bm}
\usepackage{amsmath}
\usepackage{float}

\begin{document}

\title{Zero-energy resonances in ultracold hydrogen sticking to liquid helium films of finite thickness}

\author{R. Karakhanyan}
\affiliation{P.N. Lebedev Physical Institute, 53 Leninsky prospect, 117924, Moscow, Russia}
\affiliation {Moscow Institute of Physics and Technology, 141700 Dolgoprudny, Russia}
\author{V. Nesvizhevsky}
\affiliation{Institut Laue-Langevin (ILL), 71 avenue des Martyrs, 38042, Grenoble, France}
\author{A. Semakin}
\affiliation{Department of Physics and Astronomy, University of Turku, 20014, Turku, Finland}
\author{S. Vasiliev}
\affiliation{Department of Physics and Astronomy, University of Turku, 20014, Turku, Finland}
\author{A. Voronin}
\affiliation{P.N. Lebedev Physical Institute, 53 Leninsky prospect, 117924, Moscow, Russia}

\begin{abstract}
We investigated quantum states of ultracold hydrogen atoms in a combined potential comprising the H-He film interaction in the presence of a substrate and the Earth's gravitational field. We show that the shift and width of the gravitational quantum states are determined by the complex scattering length for the H-He film/substrate potential. We demonstrate that for specific helium film thicknesses above a substrate, zero-energy resonances occur if the combined potential supports a bound state exactly at the threshold. This effect leads to a complete restructuring of the bound states spectrum. The dynamics of gravitational levels as a function of the van der Waals interaction depth---controlled by the helium film thickness---is analyzed. It reveals the critical thicknesses of $\sim$6.1~and $\sim$1.8~nm, at which resonances appear in the case of the conductive substrate. With imaginary integral operators, we incorporate non-perturbatively the inelastic effects originating from the ripplon coupling. The inelastic effects show dramatic changes in the sticking-coefficient behavior near the critical points. The enhanced sticking coefficients provide a probe for studying critical phenomena and measuring atom-surface interaction parameters with unprecedented sensitivity.
\end{abstract}


\maketitle

\section{Introduction}

The quantum states of light ultra-cold particles in the gravitational field near material surfaces have emerged as a powerful platform for precision measurements and fundamental physics tests~\cite{nesvizhevsky2002,nesvizhevsky2003,nesvizhevsky2005,nesvizhevsky2010,nesvizhevsky2010pu,voronin2011,jenke2011,jenke2014,ichikawa2014,crivelli2015,baessler2015}. These gravitational quantum states (GQS) were first observed for ultracold neutrons~\cite{nesvizhevsky2002} in a triangular potential well formed by a Fermi repulsion from the material surface and a linear attractive gravity term. For neutral atoms, a similar type of repulsion is expected, with the repulsive term replaced by a quantum reflection from the tail of the Casimir-Polder interaction of the atoms with the surface. Such reflection occurs at fairly large distances from the surface, which is especially important for antihydrogen to avoid annihilation at the matter wall. Therefore, theoretical studies of GQS for (anti)atoms were motivated mostly by the prospects of experiments with antihydrogen. Recent theoretical predictions suggest that antihydrogen GQS states above a liquid helium surface could achieve lifetimes of $\sim$1~s~\cite{voronin2005,voronin2012,dufour2013,voronin2014,dufour2014,voronin2016,crepin2017,crepin2019}. This offers unique opportunities for studying quantum mechanics in the presence of gravity, particularly precision measurements of gravitational properties, which could provide sensitive tests of the equivalence principle~\cite{nesvizhevsky2010,voronin2011}. 

In the case of normal hydrogen, the effects of the interaction with the surface at shortest distances cannot be neglected. They lead to a variety of quantum phenomena with a long history of theoretical and experimental research. A surface of superfluid helium plays a special role because of the weakest interaction with atoms and its own unique quantum properties. Using a superfluid helium film covering inner surfaces of the experimental cell allowed a first stabilization of the gas of atomic hydrogen below 1 K, which stimulated a boom in experimental research with Bose-Einstein condensation (BEC) as the main goal. The interaction of H with liquid helium is well understood and characterized (see~\cite{walraven1992,greytak1984} for reviews). The interaction potential provides a bound state with a binding energy of $\approx1.14$~K on $^4$He and $\approx$~0.3--0.4~K on $^3$He-$^4$He mixture films~\cite{safonov1998}. Unlike an antihydrogen, an H atom that was not reflected by the tail of the potential may settle into a surface-bound bound state (i.e. stick to the surface). This process is mediated by an emission of a surface excitation, a ripplon. Inelastic scattering mediated by the ripplon emission processes was found to be ineffective for temperatures below 0.5~K~\cite{kagan1984, carraro1992}. The probability of sticking decreases with the energy $E$ of the incoming atom as $\sim\sqrt{E}$, which was originally predicted nearly 90 years ago by Lennard-Jones and Devonshire \cite{lennard1936}. This behavior was confirmed in numerous experimental and theoretical studies in the temperature range below 100~mK~\cite{yu1993,hijmans1992, carraro1992,shimizu2001,friedrich2002}. Experiments revealed that the binding energy and the probability of sticking of H increase for the helium film with a reduced thickness~\cite{yu1993}, which was explained by the influence of the solid substrate under the film~\cite{hijmans1992, carraro1992}.

Observation of the GQS of hydrogen atoms is an intriguing experimental challenge. These states are extremely shallow and require atoms with effective energies in a direction normal to the surface in the nanokelvin region. Such experiments were recently proposed~\cite{voronin2011,vasiliev2019} and are pursued by the GRASIAN Collaboration~\cite{GRASIAN,killian2023,killian2024}. In this work, we analyze the behavior of the GQS and the adsorbed state(s) as a function of the substrate potential, which can be modified by changing the thickness of the helium film. We find resonance behavior at certain film thicknesses, when energy of the lowest  GQS  becomes negative and promptly decreases with a decrease of the film thickness. We analyze the role of the interactions with the ripplons of the film at these resonances and find that the sticking probability is strongly enhanced. This behavior should be measurable in future experiments.

The paper is organized as follows: in Sec.~II we discuss the potential model of H atom interaction with He film above substrate; in Sec.~III we discuss zero-energy resonances in such a potential for certain thicknesses of He film and restructuring of gravitational states spectrum due to such resonances; in Sec.~IV we study inelastic processes due to atom-ripplon coupling and behavior of a sticking coefficient in the vicinity of zero-energy resonance; in Sec.~V we discuss experimental observability of predicted phenomena; and in Sec.~VI we give conclusions.

\section{Theoretical Framework}

\subsection{Interaction potential scales}

The system of ultra-cold hydrogen atoms near liquid helium surfaces is characterized by three dramatically different energy and length scales that create unique quantum mechanical behavior:

\textit{Short-range adsorption potential}: Provides a bound state with the energy $E_b~\sim~1$~K and the spatial extent $r_b~\sim~5$~\AA~(see~\cite{walraven1992, greytak1984, mantz1979, Jochemsen1981} and references therein).

\textit{Long-range van der Waals-Casimir interaction}: Exhibits $C_4/z^4$ asymptotic behavior with the characteristic length $l_{\text{vdW}}~ =~\sqrt{2mC_4}/\hbar \sim~4$~nm for the H-He interactions and $\sim~20$~nm for the H-conducting surface interactions.

\textit{Gravitational potential}: Supports bound quantum states with the characteristic energy scale $\varepsilon_g \sim 0.6$~peV and the spatial scale $l_g \sim 6~\mu$m.

The extreme scale separation is quantified by the ratios:
\begin{equation}
\epsilon = \frac{l_g}{r_b} \sim 10^{4}, \quad \eta = \frac{\varepsilon_g}{E_b} \sim 10^{-8}.
\end{equation}

This separation enables the development of precise theoretical methods and leads to remarkable quantum phenomena, particularly when the system approaches critical conditions where new bound states appear at the threshold energy.

The concept of long-lived GQS relies on quantum reflection from the van der Waals-Casimir potential tail, which occurs at distances much smaller than the gravitational length scale $l_g$~\cite{voronin2005,berkhout1989,shimizu2001,friedrich2002,friedrich2006,druzhinina2003}. Previous work showed that the energy shift of GQS due to surface interaction can be expressed as~\cite{voronin2011}:
\begin{equation}\label{Shift E}
   \Delta E_n= m\cdot g\cdot a,
\end{equation}
where $g$ is the gravitational acceleration and $a$ is the complex scattering length characterizing the H-He surface potential.

This expression reveals that all GQS acquire identical shifts to the first order in the small parameter $a/l_g$, leaving transition frequencies between GQS essentially unaffected by the surface interaction. This property enables precision measurements of gravitational effects in quantum systems~\cite{voronin2011,killian2023,killian2024}.

However, this perturbative treatment breaks down if the system approaches a critical regime where $a \to \infty$. This occurs if the thickness of the helium film is adjusted such that the combined H-surface potential superposed with an H-substrate potential~\cite{hijmans1992,tiesinga1990} supports a new bound state exactly at zero energy, a zero-energy resonance. In this critical regime, the gravitational state spectrum undergoes complete restructuring, which we analyze in detail in this work.

\subsection{Combined potential structure}

The total effective potential for a hydrogen atom above a helium-coated substrate is assumed to take the form:
\begin{equation}\label{Pot}
V(z) = V^{\text{He}}(z) + V^{\text{sub}}(z,d) + mgz,
\end{equation}
where $V^{\text{He}}(z)$ represents the H-He surface interaction, $V^{\text{sub}}(z,d)$ describes the H-substrate interaction modified by the helium film of thickness $d$, and $mgz$ is the gravitational potential.

The H-He surface interaction combines short-range repulsion with long-range van der Waals attraction~\cite{walraven1992}. We model this using an effective potential of the form:
\begin{equation} \label{Vad}
V_{\text{ads}}(z) =\begin{cases}
D_e[e^{-2\beta(z-z_e)} - 2e^{-\beta(z-z_e)}],& z<z_c\\
V_{\text{vdW}}^{\text{He}}(z), & z\ge z_c
\end{cases}.
\end{equation}

The parameters are determined from experimental measurements of the H-He binding energy~\cite{mantz1979, morrow1981}:
\begin{table}[h]
\centering
\caption{Parameters of the H-He adsorption potential.}
\begin{tabular}{c|c|c}
\hline
Parameter & Value & Units \\
\hline
$D_e$ & 5.14 & K \\
$\beta$ & 0.52 & \AA$^{-1}$ \\
$z_e$ & 4.2 & \AA \\
$z_c$ & 13.0 & \AA \\
\hline
\end{tabular}
\label{tab:parameters}
\end{table}

The van der Waals interaction exhibits the standard asymptotic forms:
\begin{equation}\label{vdW}
V_{\text{vdW}}^{\text{He,sub}}(z) = \begin{cases}
-\frac{C_3}{z^3}, & z \ll \lambda_A \\
-\frac{C_4}{z^4}, & z \gg \lambda_A
\end{cases},
\end{equation}
with $\lambda_A \approx 121$~nm being the wavelength of the hydrogen 1S$\to$2P transition. The van der Waals constants (in atomic units) are determined from theoretical calculations and experimental data~\cite{carraro1992,goldman1986}:
\begin{align}
C_3^{\text{He}} &= 4.5\times 10^{-3}, \quad C_4^{\text{He}} = 1.55,\\
C_3^{\text{cond}} &= 0.25, \quad C_4^{\text{cond}} = 73.62\\
C_3^{\text{sil}} &= 0.10, \quad C_4^{\text{sil}} = 50.28
\end{align}
for H-He and H-conducting surface and H-Silicon interactions, respectively.

The contribution of the substrate through a helium film of thickness $d$ is:
\begin{equation}
V^{\text{sub}}(z,d) = V_{\text{vdW}}^{\text{sub}}(z+d) - V_{\text{vdW}}^{\text{He}}(z+d).
\end{equation}

For distances $z \gg d$, the potential exhibits asymptotic behavior:
\begin{equation}\label{Pot-as}
V_{\text{vdW}}(z)\sim -\frac{C_4^{\text{cond}}+C_4^{\text{He}}}{z^4}.
\end{equation}

Since $C_4^{\text{sub}} \gg C_4^{\text{He}}$ for conducting substrates, long-range behavior and quantum reflection properties are mainly determined by the presence of the substrate, at least for $d < l_{\text{vdW}}$~\cite{dufour2013,dufour2014}.

\begin{figure}[h]
\centering
\includegraphics[scale=0.4]{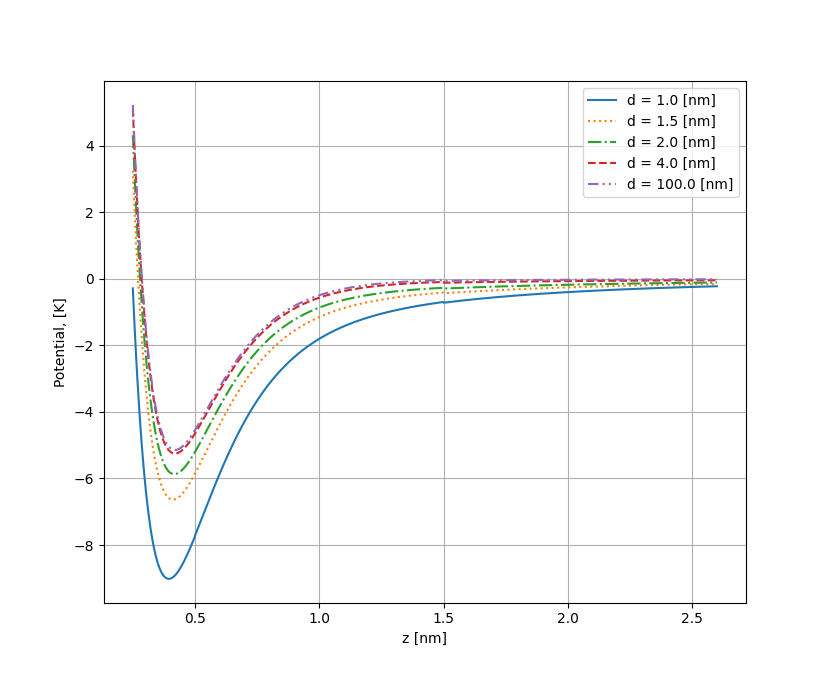}
\caption{Total potential $V(z)$ as a function of height $z$ above the surface for different helium film thicknesses $d$ above a conducting substrate. The gravitational component $mgz$ dominates at large distances, while the van der Waals interaction determines the short-range behavior.}
\label{fig:potential}
\end{figure}

\subsection{Boundary condition for gravitational states}

The extreme scale separation between gravitational and surface potentials allows us to treat their coupling through a boundary condition applied at an intermediate distance $l_{\text{vdW}} \ll z_0 \ll l_g$~\cite{voronin2011,voronin2014,voronin2016}. This approach connects the wave function behavior in the surface interaction region with that in the gravitational field region.

For the surface interaction region, the asymptotic wave function behavior at $z_0 \gg l_{\text{vdW}}$ takes the form:
\begin{equation}\label{Left}
\left.\frac{\psi}{d\psi/dz}\right|_{z_0} = z_0-a,
\end{equation}
where $a$ is the complex scattering length determined entirely by the surface potential.

In the gravitational region, the wave function is described by the Airy function:
\begin{equation}\label{GravWF}
\psi_n(z) \sim \text{Ai}\left(\frac{z}{l_g} - \lambda_n\right),
\end{equation}
where $l_g = (\hbar^2/2m^2g)^{1/3}$ is the gravitational length.

Matching these solutions yields the eigenvalue equation:
\begin{equation}\label{EQEgrav}
\frac{z_0-a}{l_g}= \frac{\text{Ai}\left(\frac{z_0}{l_g} - \lambda_n\right)}{\text{Ai}'\left(\frac{z_0}{l_g} - \lambda_n\right)}.
\end{equation}

For $a \ll l_g$, this reduces to the perturbative result of Eq.~(\ref{Shift E}), where all gravitational states acquire the same energy shift $\Delta E = mga$.

Calculations of the scattering length give
\begin{equation}\label{aHbarHe}
a_{H} = -34.98 - 44.84i \text{ a.u.}
\end{equation}
for antihydrogen on bulk liquid He  and 
\begin{equation}\label{aHHe}
a_{H} = -18.43 \text{ a.u.}
\end{equation}
for hydrogen on bulk liquid  He ~\cite{crepin2017}.

These values lead to small, uniform energy shifts for all GQS, which confirms the validity of the perturbative approach for typical conditions.

\section{Zero-energy resonances and spectrum restructuring}

\subsection{Critical film thicknesses}

The situation changes dramatically if the helium film thickness reaches critical values at which the surface potential supports a new bound state exactly at zero energy. At these points, the scattering length diverges according to the following:
\begin{equation}
|a(E)| \sim \frac{\hbar}{\sqrt{2m|E_b|}} \rightarrow \infty \text{ as } E_b \rightarrow 0,
\end{equation}
where $E_b$ is the binding energy of the near-threshold state.

\begin{figure}[h]
\centering
\includegraphics[width=1\linewidth]{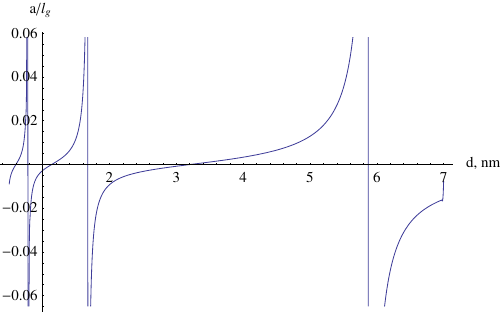}
\caption{Scattering length $a$ as a function of helium film thickness $d$. Sharp peaks indicate zero-energy resonances at which $|a| \to \infty$.}
\label{fig:scattering}
\end{figure}

Our numerical calculations reveal two critical film thicknesses for hydrogen above the conducting substrates (see Fig.~\ref{fig:scattering}):
\begin{align}
d_1^{\text{He}} &= 6.1 \text{ nm}\\
d_2^{\text{He}} &= 1.8 \text{ nm}.
\end{align}

For the critical case $a \gg l_g$, Eq.~(\ref{EQEgrav}) can be rewritten as:
\begin{equation}\label{EQEgrav1}
\text{Ai}'\left(\frac{z_0}{l_g} - \lambda_n\right)= \frac{l_g \text{Ai}\left(\frac{z_0}{l_g} - \lambda_n\right)}{z_0-a}.
\end{equation}

Since this equation must be independent of the matching point $z_0$, we obtain:
\begin{equation}\label{EQEgrav2}
\text{Ai}'\left( - \lambda_n\right)= -\frac{l_g\text{Ai}\left( - \lambda_n\right)}{a}.
\end{equation}

In the limit $a/l_g \to \infty$, the boundary condition becomes:
\begin{equation}
\text{Ai}'(-\Lambda_n) = 0.
\end{equation}

This represents a complete restructuring of the gravitational spectrum; now, eigenvalues are determined by zeros of the Airy function derivative rather than the function itself.

\begin{table}[h]
\centering
\caption{Comparison of gravitational eigenvalues in regular and threshold resonance cases.}
\begin{tabular}{c|c|c|c}
\hline
Level & Threshold case, $\Lambda_n$ &  Regular case, $\lambda_n$ & Relative shift \\
\hline
1 & 1.407 & 2.338 & -39.8\% \\
2 & 2.461 & 4.088 & -39.8\% \\
3 & 3.324 & 5.521 & -39.8\% \\
4 & 4.086 & 6.787 & -39.8\% \\
\hline
\end{tabular}
\label{tab:restructuring}
\end{table}

Near resonance, the energy shift can be expressed as:
\begin{equation}\label{Shift E*}
\Delta E_n^*=\frac{mg l^2_g}{a \Lambda_n}.   
\end{equation}

\subsection{Evolution of the bound state spectrum}

\begin{figure}[h]
\centering
\includegraphics[width=1\linewidth]{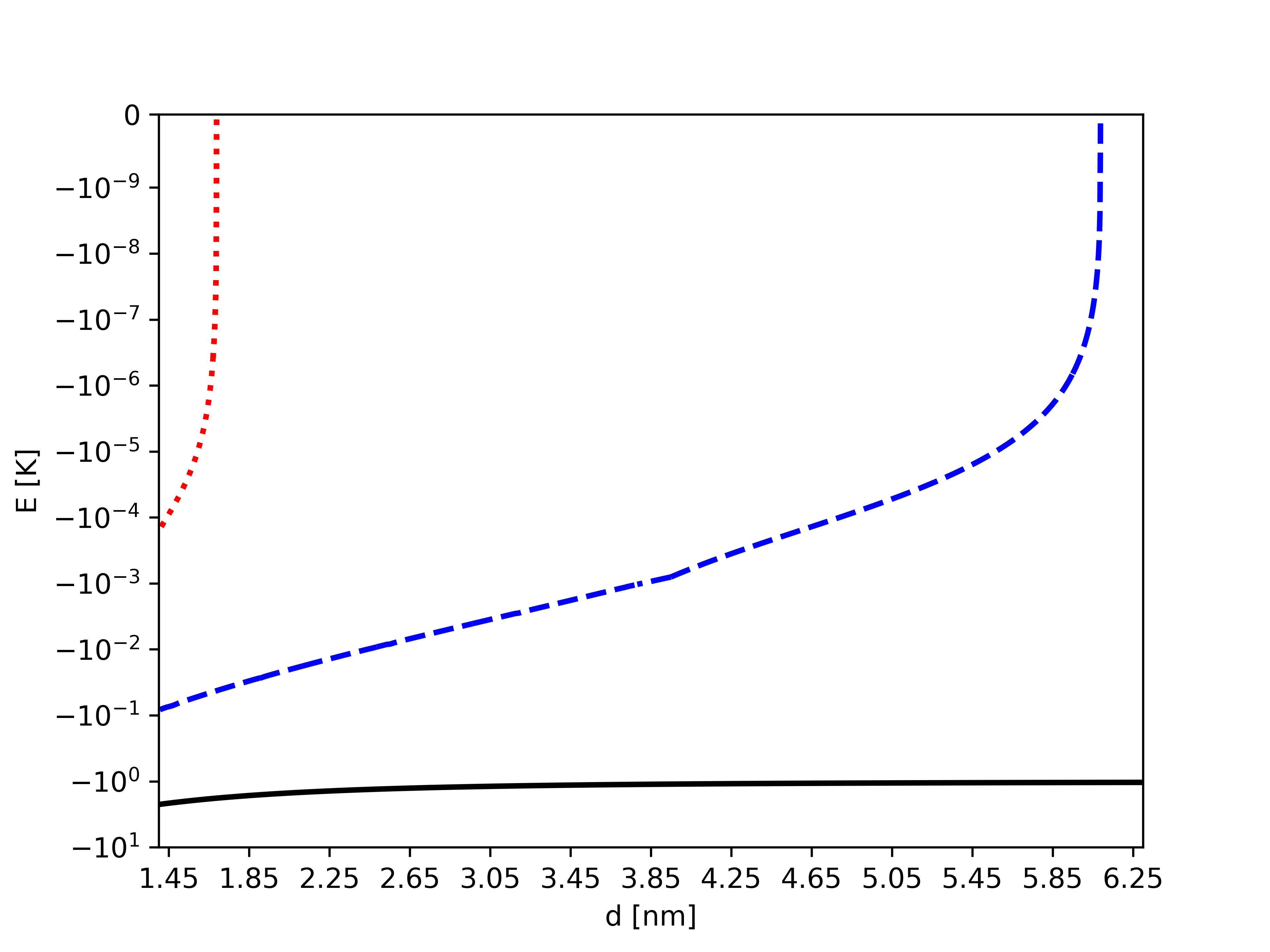}
\caption{Energy evolution of the adsorption state and two lowest GQS versus helium film thickness on logarithmic scale. Sharp discontinuities mark critical thicknesses at which spectrum restructuring occurs. Black curve- ground adsorption state, blue- first excited state, red- second excited state.}
\label{fig:levels_total}
\end{figure}

\begin{figure}[h]
\centering
\includegraphics[width=1\linewidth]{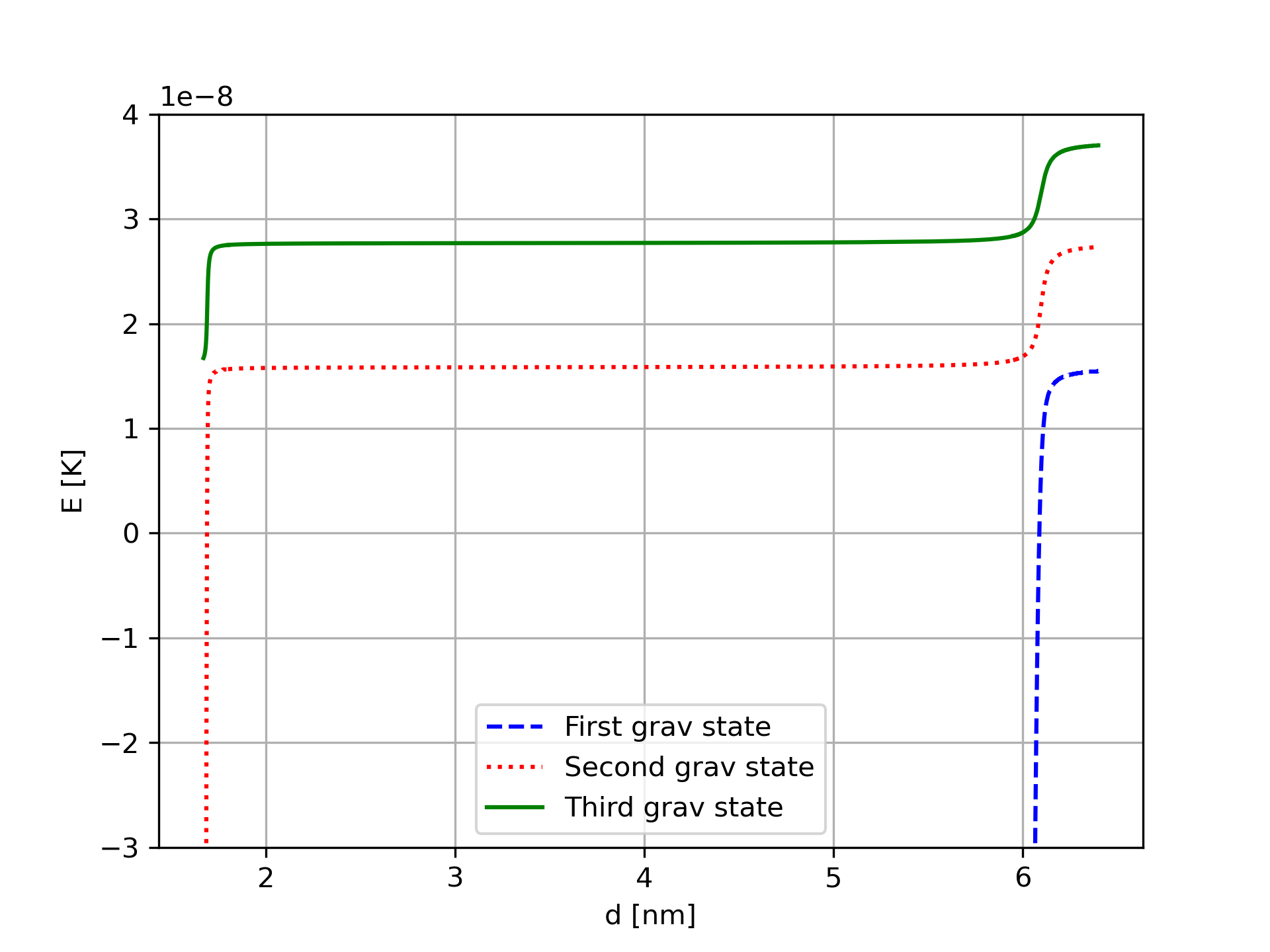}
\caption{Detailed view of the three lowest GQS energies near the critical film thicknesses at $d_1 = 6.1$~nm and $d_2=1.8$~nm.}
\label{fig:levels_critical}
\end{figure}

We calculated the evolution of both the adsorption and gravitational states as functions of the helium film thickness. The results reveal a pattern of level-avoidance crossings near certain critical points $d=6.1$ nm and $d=1.8$ nm as one can see in Fig.\ref{fig:levels_total} and Fig.\ref{fig:levels_critical}.

The energy of the adsorption state remains approximately constant at $\sim 1$~K until the film thickness decreases below $\sim 3$~nm, where the attraction of the substrate begins to significantly deepen the binding energy.

In contrast, GQS exhibit dramatic behavior near the critical thicknesses. Below the first critical thickness $d_1^{\text{He}} = 6.1$~nm, the lowest GQS rapidly transitions to negative energies, effectively "diving" into the surface potential well. Furthermore, higher GQS shift downward to fill the vacated energy regions as shown on Fig.~\ref{fig:levels_critical}.

At the second critical thickness $d_2^{\text{He}} = 1.8$~nm, the system supports three negative energy states, with the binding energy of the second adsorption state reaching 0.04~K.

\section{Inelastic effects and nonperturbative treatment}

\subsection{Breakdown of perturbation theory near resonance}

Helium surfaces support quantum excitations (ripplons) that couple to hydrogen atoms, leading to inelastic scattering processes~\cite{kagan1984,salonen1982,goldman1986,berkhout1993}. Traditionally, such inelastic scattering, in particular sticking, has been treated within the perturbation approach due to the weak coupling of the $H$ atom motion with surface excitations with typical amplitudes of a few \AA.

However, in in near zero-energy resonances, the enhanced scattering length dramatically amplifies the coupling between the atoms and surface excitations, invalidating the perturbative approach. We develop a non-perturbative treatment by incorporating ripplon interactions through complex integral operators that preserve the physical unitarity constraints.
To restore physical behavior, we introduce an effective imaginary operator in the Schrödinger equation for the elastic channel. We choose this operator in such a way that perturbation theory results are restored far from resonance ~\cite{zimmerman1983,hijmans1992}:
\begin{equation}\label{Schr1}
\left[-\frac{\hbar^2}{2m}\frac{d^2}{dz^2}+V(z)-i\hat{W} -E\right]\psi_s(z)=0.
\end{equation}

The effective inelastic operator has the separable form:
\begin{equation}\label{AbsorptionPotential}
\hat{W}=h^2_{q_a} \left|\frac{\partial U_{q_a}(z)}{\partial z} \psi_b \right\rangle \left\langle \psi_b \frac{\partial U_{q_a}(z)}{\partial z} \right|, 
\end{equation}
where the coupling matrix element uses the Morse-type expression~\cite{zimmerman1983}:
\begin{align}\label{Uq}
&\frac{\partial U_{q_a}(z)}{\partial z}=2\beta D_e\left[\frac{e^{-2\beta(z-z_e)\sqrt{1+q_a^2/(4\beta^2)}}}{\sqrt{1+q_a^2/(4\beta^2)}} \right.\\
&\left.- 2\frac{e^{-\beta(z-z_e)\sqrt{1+q_a^2/\beta^2}}}{\sqrt{1+q_a^2/\beta^2}}\right]\nonumber.
\end{align}

The separable form reflects the treatment of a single dominant inelastic channel, justified by the energy hierarchy of the problem. The above form of the effective inelastic operator is chosen to reproduce perturbative results for the sticking coefficient in the case of the bulk He surface.

\subsection{Sticking probability}

The small momentum expansion of sticking probability yields the following expression:
\begin{equation}\label{Res}
|S(E)|^2=1-\frac{(p-a_1/b_0)^2}{(p+a_1/b_0)^2}.
\end{equation}
Here, the term $a_1$ is an even function of momentum associated with the account of the imaginary inelastic operator $-i\hat{W}$, while $b_0$ is an even function of momentum, associated with the interaction of the atom-flat surface.
This shows a maximum at $p = a_1/b_0$ and restores the proper behavior $\sqrt{E}$ at very low energies, resolving the unphysical divergence~\cite{doyle1991,pasquini2004}.

\begin{figure}[h]
\centering
\includegraphics[width=1.0\linewidth]{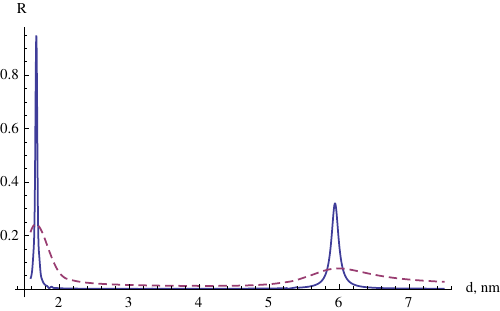}
\caption{Sticking coefficient versus He film thickness on the ideal conductor substrate for incident He atoms with the energy $10^{-5}$~K (dashed line) and $10^{-7}$~K (solid line), showing dramatic enhancement near the critical thicknesses.}
\label{fig:Sticjd}
\end{figure}

\begin{figure}[h]
\centering
\includegraphics[width=1.0\linewidth]{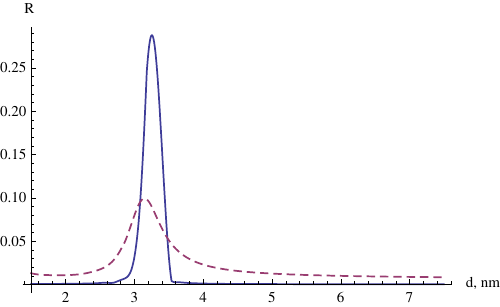}
\caption{Sticking coefficient versus He film thickness on a silicon substrate for incident H atoms with the energy $10^{-5}$~K (dashed line) and $10^{-7}$~K (solid line), showing dramatic enhancement near the critical thicknesses.}
\label{fig:SticSilica}
\end{figure}

We demonstrate  the dependence of the sticking probability on the thickness of the $He$ film in the case of the ideal conducting substrate in Fig.~\ref{fig:Sticjd} and the silicon substrate in Fig.~\ref{fig:SticSilica}. These plots show clear resonant properties in the vicinity of the critical thickness $d=6.1$ nm.

\begin{figure}[h]
\centering
\includegraphics[width=1.0\linewidth]{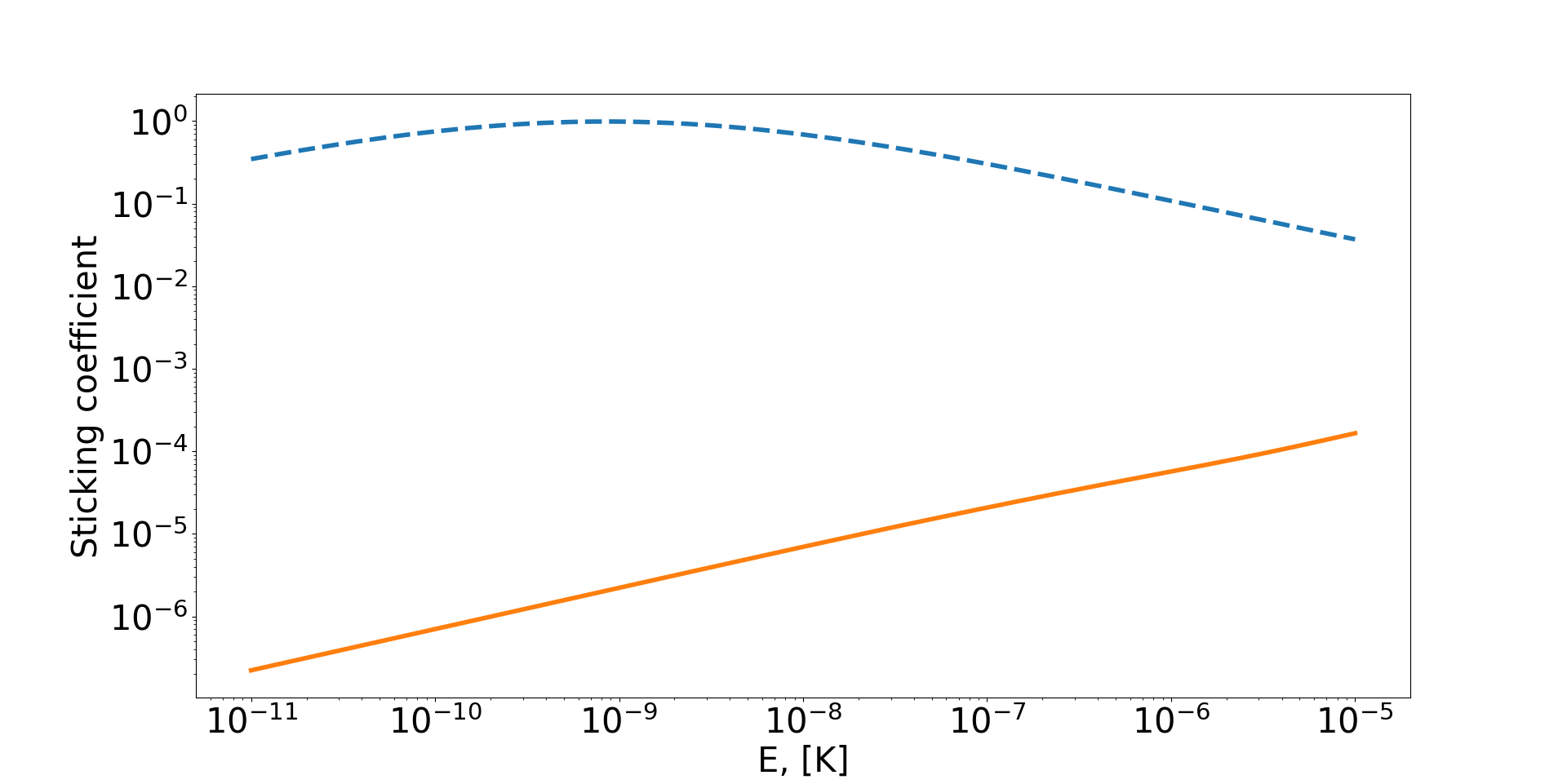}
\caption{Sticking probability versus incident energy for the near resonance thickness of $6.1$ nm (dashed blue line) and bulk liquid helium (yellow solid line), demonstrating clear departure from the universal $\sqrt{E}$ behavior for the resonant case.}
\label{fig:SticjE}
\end{figure}

We show the energy dependence of the sticking probability in the vicinity of the critical thickness of the $He$ film above the ideal conducting substrate in Fig.\ref{fig:SticjE}. The probability of sticking has a clear resonance profile in this case according to the expression (\ref{Res}). It differs radically from standard quantum reflection $\sqrt{E}$ behavior in non-resonant case.

\begin{figure}[h]
\centering
\includegraphics[width=1\linewidth]{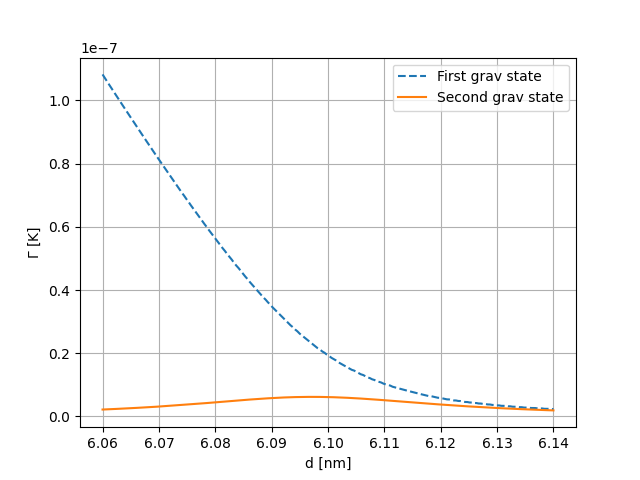}
\caption{Gravitational state width $\Gamma_n$ versus helium film thickness, showing dramatic increases of the lowest GQS width near critical point at which the GQS transforms into a localized adsorption state. Dashed line- first state, solid line- second state}
\label{fig:GammaD}
\end{figure}

\begin{figure}[h]
\centering
\includegraphics[width=1\linewidth]{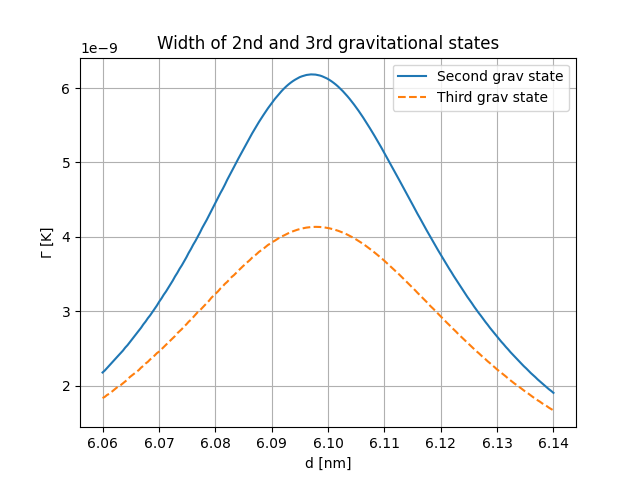}
\caption{Excited GQS widths $\Gamma_n$ versus helium film thickness near the first critical point. Solid line- second state, dashed line- third state}
\label{fig:GammaD2}
\end{figure}

Inelastic interactions lead to finite state lifetimes:
\begin{equation}
\tau_n = \frac{\hbar}{\Gamma_n},
\end{equation}
where the width is:
\begin{equation}
\Gamma_n = 2\int_0^\infty \psi^*_n(z)W(z,z')\psi_n(z')dzdz'.
\end{equation}
 The corresponding dependence is demonstrated in Figs.~\ref{fig:GammaD} and ~\ref{fig:GammaD2}.

The dramatic enhancement in sticking coefficients and GQS width near critical thicknesses provides a probe for studying critical phenomena and measuring atom-surface interactions with unprecedented sensitivity~\cite{pasquini2006,vasiliev2022}.

\section{Experimental observability and applications}

The critical He film thicknesses of 6.1~nm and 1.8~nm are within the typical experimental control ranges. The properties of helium films up to submonolayer coverages (1 monolayer $\approx 0.36$~nm) were studied in numerous experiments (see~\cite{hallock1995} for a review). However, not only do the adsorbed hydrogen properties strongly depend on the structure of the substrate, but the same concerns a few first helium layers above the solid substrate. For most solid substrates, it was found that the first 1-2 layers of helium are in the solid state~\cite{bashkin1995, cheng1992}, with the only exception of a solid molecular hydrogen. Helium has so weak binding potential on H$_2$ that the adsorbed phase remains in a superfluid state even in the sub-monolayer region of coverages~\cite{shirron1991}. In general, few monolayer helium films near any solid substrate have a layered structure~\cite{cheng1992, bashkin1995,krotscheck2008} that is not taken into account in our model. Therefore, we restricted our consideration to the range of thicknesses above 1.2 nm, at which the helium film is liquid and has a smooth density profile. 

The predicted spectrum restructuring could be observable through several experimental signatures:

\textbf{Adsorption energy}: The adsorption energy was measured for the thickness of the film down to 1 nm by an indirect method, measuring the temperature dependence of the two-body recombination rate in the adsorbed phase~\cite{godfried1985}. An increase in adsorption energy was observed for the thickness of the film below 2~nm, qualitatively matching the results in Fig.~\ref{fig:levels_total} for the ground state. The possible emergence of the second bound state could not be detected by this method. Unfortunately, direct and more accurate experiments with the adsorbed phase, in which the adsorbed atoms were detected by magnetic resonance~\cite{shinkoda1991,pollack1992,jarvinen2005}, were performed only for thick saturated films. In the ESR experiments~\cite{shinkoda1991,jarvinen2005}, emergence of the extra bound state would be easily detected as the appearance of an extra line in the spectrum.

\textbf{Magnetic resonance (wall) shift}: The interaction of the H atoms with the substrate also leads to a modification of the hyperfine constant of the adsorbed atom, known as a wall shift. This shift is the main source of uncertainties in the frequency of the hydrogen maser. For H adsorbed on saturated helium films it was measured by hyperfine~\cite{morrow1981}, NMR~\cite{pollack1992} and electron-nuclear double resonance spectroscopy~\cite{ahokas2007}. In the latter work, the wall shift for the lowest bound state of H on helium $\Delta \nu=-45.58(4)$ kHz was measured. This implies a fractional decrease $3.2\cdot10^{-5}$ of the hyperfine constant measured with the accuracy of $\approx10^{-4}$. Clearly, the presence of the extra bound states, located further from the surface, can be easily detected by this method. Much better accuracy can be achieved using a cryogenic version of the hydrogen maser operating with surface-adsorbed atoms~\cite{maan1993}.

\textbf{Sticking probability}: The dramatic change in energy dependence from $\propto \sqrt{E}$ to resonant profiles [Eq.~(\ref{Res})] provides possible experimental signatures that can be measured using techniques developed for ultra-cold hydrogen research~\cite{yu1993,morrow1981,berkhout1993}. The only measurement of the probability of sticking to the thin films down to 0.5~nm was done for H at energies $\geq300~\mu$K~\cite{yu1993}. The effects of the substrate were well seen in an increase in the sticking probability at reduced film thickness. Scattering resonances were not observed. Measurements were made with a porous sinter surface with non-uniform thickness, which averaged possible resonance behavior. Obviously, such measurements should be performed with a well-characterized and smooth surface. The surface of solid molecular hydrogen is of special interest because of the weakest van der Waals interaction. Such coverages are usually produced in any experiment with atomic hydrogen as a result of a recombination of atoms with subsequent build-up of the molecular layer under the helium film.

\textbf{Gravitational state spectroscopy and interferometry}: 

Restructuring of the GQS spectrum, for which energy levels shift by almost 40\%, can be measured in flow experiments studying GQS. One such experiment, in particular, is currently being conducted in Vienna as part of the GRASIAN project~\cite{killian2023,killian2024}. Its characteristic sensitivity/precision is at least one percent. This requires replacing the vacuum chamber with a mirror and absorber at room temperature with a cryostat with the ability to coat the existing mirror with a He film and control its thickness. It is important that this method allows for the separation of the actual quantum reflection of atoms from a surface without contact with the surface and contact interaction as well as the measurement of the probabilities of H sticking to the surface.

Much more precise spectroscopic and interferometric studies of GQS can be performed with ultra-cold H atoms after deep cooling in the $\mu$K regime in a magnetic trap. This will allow for a substantial increase in the lifetime of the H GQS, a decrease in the velocities of H, and an increase in the density of the GQSs in the phase space. This can be achieved in the coming years in the experiments currently being prepared in Turku, within the framework of the GRASIAN project~\cite{vasiliev2019,vasiliev2020,vasiliev2022}.

\textbf{Lifetime measurements}: The increase in GQS widths by several orders of magnitude near critical points provides another experimental observable through measurements of GQS lifetimes.

All of the above analysis was performed under the assumption of a smooth superfluid He surface perturbed by ripplon excitations. The relatively simple model of atom-ripplon interaction, including the correlation length of excitations on a helium surface, could be tested with much higher sensitivity under the near-resonance conditions discussed here.

\section{Conclusions}

We demonstrated that the helium film thickness provides sensitive control over quantum-state spectra of ultra-cold hydrogen atoms, enabling transitions between gravitational and adsorbed state regimes. For conductive substrates, at critical thicknesses of $\sim$~6.1~nm and $\sim$~1.8~nm, zero-energy resonances trigger complete spectrum restructuring with energy shifts exceeding 39~\%, unifying gravitational and surface-bound states into a single quantum mechanical framework. 

A possibility of the emergence of the second bound state and related scattering resonance was mentioned in theoretical works by Carraro and Cole~\cite{carraro1992,carraro1998} and later by Krotscheck and Zillich~\cite{krotscheck2008}. In the latter work, the sticking resonance was predicted to occur at the thickness of the helium film in the range between 5 and 2~nm. This is in qualitative agreement with the result of this work considering a somewhat different form of the interaction potential. Unfortunately, no quantitative estimates were provided in the aforementioned theoretical considerations. The effects of gravity at the tail of the interaction potential and existence of the gravitational bound states are considered here for the first time. 

The theoretical framework developed here successfully describes the extreme scale separation ($\epsilon~\sim~10^4$, $\eta~\sim~10^{-8}$) through the boundary condition methods that relate surface and gravitational physics~\cite{voronin2016,nesvizhevsky2019}. The breakdown of perturbation theory near zero-energy resonances is resolved through nonperturbative treatment of inelastic ripplon coupling, preserving physical behavior and predicting finite lifetimes for restructured states.

The enhanced sticking coefficients and modified energy dependencies near critical points provide sensitive probes of atom-surface interactions and critical quantum phenomena. The transition from standard $\sqrt{E}$ energy dependence to resonant profiles with finite maxima [Eq.~(\ref{Res})] offers unprecedented sensitivity for studying fundamental atom-surface interactions.

The predicted phenomena occur at film thicknesses and energy scales accessible to current experimental techniques, making immediate verification possible~\cite{vasiliev2022,killian2024}.

This work establishes ultracold hydrogen on engineered helium surfaces as a powerful platform for studying critical quantum phenomena and conducting precision tests of fundamental physics. Future experimental verification of these predictions could open new research directions in quantum metrology, precision tests of gravitational interactions with quantum matter, and studies of quantum phase transitions in atom-surface systems. 


\begin{acknowledgments}
We acknowledge helpful discussions with colleagues in the gravitational quantum states community and support from relevant funding agencies. Special thanks to the GRASIAN collaboration for motivating this theoretical investigation.
\end{acknowledgments}

\bibliography{refs}

\end{document}